\documentclass[12pt]{JHEP3}

\usepackage{epsf,epsfig,psfig}

\def\spose#1{\hbox to 0pt{#1\hss}}
\def\ltapprox{\mathrel{\spose{\lower 3pt\hbox{$\mathchar"218$}}
 \raise 2.0pt\hbox{$\mathchar"13C$}}}
\def\gtapprox{\mathrel{\spose{\lower 3pt\hbox{$\mathchar"218$}}
 \raise 2.0pt\hbox{$\mathchar"13E$}}}

\title{
Topological susceptibility of 
SU($N$) gauge theories at finite temperature
}

\author{Luigi Del Debbio \\
CERN, Department of Physics, TH Division, CH-1211 Geneva 23\\
	E-mail: \email{Luigi.Del.Debbio@cern.ch} 
} 

\author{Haralambos Panagopoulos \\ 
        Department of Physics, University of Cyprus\\
        Lefkosia, CY-1678, Cyprus\\ 
	E-mail: \email{haris@ucy.ac.cy} 
}

\author{Ettore Vicari \\ 
	Dipartimento di Fisica dell'Universit\`a 
	di Pisa and I.N.F.N. \\
        Via Buonarroti 2, I-56127 Pisa, Italy \\ 
	E-mail: \email{vicari@df.unipi.it} 
}

\abstract{ We investigate the large-$N$ behavior of the topological
 susceptibility $\chi$ in four-dimensional SU($N$) gauge theories at
 finite temperature, and in particular across the finite-temperature
 transition at $T_c$.  For this purpose, we consider the lattice
 formulation of the SU($N$) gauge theories and perform Monte Carlo
 simulations for $N=4,6$.  The results indicate that $\chi$ has a
 nonvanishing large-$N$ limit for $T<T_c$, as at $T=0$, and that the
 topological properties remain substantially unchanged in the
 low-temperature phase.  On the other hand, above the deconfinement
 phase transition, $\chi$ shows a large suppression.  The comparison
 between the data for $N=4$ and $N=6$ hints at a vanishing large-$N$
 limit for $T>T_c$.  } 


\preprint{CERN-PH-TH/2004-111}

\begin{document}

The pattern of chiral symmetry breaking for QCD with $N_f$ light
flavors at zero temperature is well understood. The symmetry group of
the classical lagrangian, $\mathrm{ U}(N_f)_\mathrm{L} \otimes
\mathrm{ U}(N_f)_\mathrm{R}$, is broken both by the anomaly and
spontaneously. The spontaneous breaking of the axial subgroup
$\mathrm{ SU}(N_f)_A$ yields $N_f^2-1$ Goldstone bosons, while the
anomalous breaking of the $\mathrm{ U}(1)_A$ symmetry explains the
heavier flavor-singlet state observed in the hadronic spectrum. (See
e.g. Ref.~\cite{Weinberg-book}.)

In the large-$N$ limit~\cite{Hooft-74}, where $N$ is the number of colors, the
mass of the singlet, $m_{\eta^\prime}$, is related to the topological
susceptibility of the pure gauge theory, $\chi$, through the
well-known Witten--Veneziano (WV)
formula~\cite{Witten:1979vv,Veneziano:1979ec}:
\begin{equation}
F^2_\pi m^2_{\eta^\prime}=2 N_f \chi.
\label{eq:WV}
\end{equation}
It is particularly interesting to study the topological susceptibility
as $N$ is varied, since, at fixed number of flavors, $1/N$ can be
identified with the explicit symmetry-breaking parameter for the
$\mathrm{U}(1)_A$ symmetry. Hence QCD is expected to recover the full
$\mathrm{ U}(N_f)_\mathrm{L} \otimes \mathrm{ U}(N_f)_\mathrm{R}$
chiral symmetry as $N\to\infty$. It is then possible to show that,
under very general assumptions, this chiral symmetry is spontaneously
broken down to the vector subgroup $\mathrm{ U}(N_f)_V$, yielding
$N_f^2$ massless Goldstone bosons~\cite{Coleman:1980mx}. In this
limit, if the topological susceptibility of the pure gauge theory does
not vanish, the $\eta^\prime$ acquires a mass $m^2_{\eta^\prime} \sim
1/N$, i.e. it becomes a Goldstone boson whose mass squared vanishes
linearly in the symmetry-breaking parameter $1/N$ as the anomaly is
suppressed, see e.g.~\cite{Witten:1980sp}. At zero temperature,
numerical evidence from lattice simulations in favor of a
non-vanishing large-$N$ limit of $\chi$, with $1/N^2$ power-law
corrections, has only been obtained
recently~\cite{DelDebbio:2002xa,Lucini:2001ej}.

As the temperature is increased, the $\mathrm{SU}(N)_A$ chiral
symmetry is restored at a critical temperature $T_c$.  The nature of
this phase transition, besides its own theoretical interest,
determines the dynamics of the transition from hadronic matter to a
quark--gluon plasma, which is expected to take place, e.g. in
heavy-ion collisions. In this respect, the effective breaking of the
U(1)$_A$ symmetry at finite temperature, and in particular around the
transition at $T_c$, is of particular interest. Indeed, in the case of
two light flavors, the transition may be continuous (for massless
quarks) and in the O(4) universality class only for a sufficiently
large breaking of the U(1)$_A$ symmetry around $T_c$
\cite{Pisarski:ms,Butti:2003nu}.  At finite temperature the anomaly,
considered as an equation between operators, remains
unchanged~\cite{Itoyama:up}, but its physical effect might be
drastically different because the quantities that enter the WV formula
have their own temperature dependence.  In order to verify that the WV
mechanism is still at work at finite temperature and in the
low-temperature phase, we study the behaviour of the topological
susceptibility $\chi$ as the critical temperature is approached from
below, verifying that $\chi$ has a nonvanishing large-$N$ limit,
similarly to what happens at $T=0$.  As a consequence,
Eq.~(\ref{eq:WV}) is expected to hold up to $T_c$.  The transition
between the high- and low-temperature regimes is not fully
understood. At high temperatures, where instanton calculus is
reliable~\cite{Gross:1980br}, a rather different scenario emerges.
Concerning the topological properties, an interesting hypothesis has
been put forward in Ref.~\cite{KPT-98}: at large $N$, configurations
with non-trivial topological charge are exponentially suppressed in
the high-temperature phase, i.e. behave as $e^{-N}$, so that the
topological susceptibility gets rapidly suppressed in the large-$N$
limit.  In order to shed some light on this issue, we also present
results from simulations above $T_c$.

The behavior of the topological susceptibility across the transition
has already been investigated for $N=2$ and $N=3$ in a number of
works, see e.g.
Refs.~\cite{Hoek:1986hq,Hoek:1986nd,Teper:ed,ADD-97,Gattringer:2002mr,Lucini:2004yh}.
The main focus of this work is on the large-$N$ behavior. For this
purpose we report numerical results for SU($N$) gauge theories with
$N=4,6$.  Some results for $N>3$ have recently been reported in
Ref.~\cite{Lucini:2004yh}.

The behavior of the topological susceptibility at the
finite-temperature deconfinement phase transition is studied in detail
for SU($N$) gauge theories with $N=4,6$, exploiting their lattice
formulation given by the action
\begin{equation}
S = - N \gamma \sum_{x,\mu>\nu} {\rm Tr} \left[
U_\mu(x) U_\nu(x+\mu) U_\mu^\dagger(x+\nu) U_\nu^\dagger(x) 
+ {\rm h.c.}\right],
\label{eq:wilsonac}
\end{equation}
where $U_\mu(x)\in$ SU($N$) are link variables. In order to study the
large-$N$ limit, it is convenient to replace the more familiar
coupling $\beta$ by $\gamma=\beta/2 N^2$, which is the inverse of the
't~Hooft coupling $\lambda=g^2 N$.  In order to study the theory at
finite temperature, we perform Monte Carlo simulation on asymmetric
lattices.  The gauge configurations are generated using a mixture of
microcanonical and heat-bath updating algorithms (see
Ref.~\cite{DelDebbio:2001kz,DelDebbio:2001sj} for details).  In our
simulations we consider different time extensions $L_t=6,8$ and
constant aspect ratio $L_s/L_t=4$.  These values of $L_t$ and $L_s$
should be sufficiently large to obtain results with small scaling and
finite-size corrections.  The physical temperature is given as a
function of the lattice spacing and of the lattice time extension,
$T=1/a(\gamma) L_t$.  Previous investigations have already shown the
existence of a finite-temperature phase transition, which is first
order for $N\geq 3$, see e.g. Ref.~\cite{Lucini:2003zr}. For each
value of $L_t$, $\gamma$ is tuned so as to explore an interval around
the critical temperature $T_c$; the corresponding critical value of
the coupling is denoted by $\gamma_c$. In this work the critical
couplings obtained in Ref.~\cite{Lucini:2003zr} are used,
i.e. $\gamma_c(L_t=6)=0.33717(2)$ and $\gamma_c(L_t=8)=0.34640(7)$ for
$N=4$, and $\gamma_c(L_t=6)=0.34508(5)$ for $N=6$. When
zero-temperature quantities are needed, we perform the corresponding
computation on symmetric lattices at the required values of the
coupling. A summary of our runs is presented in Tables~\ref{tab4}
and~\ref{tab6}, for $N=4$ and $N=6$ respectively.

\TABLE[ht]{
\caption{Finite-temperature data for the SU(4) gauge theory.
$\sigma_{T=0}$ and $\chi_{\,T=0}$ are respectively
the string tension and the topological
susceptibility at $T=0$ obtained using symmetric lattices.
Data marked by an asterisk are subject to an uncontrolled
systematic error due to the fact that no clear plateau
was observed in the cooling procedure to determine $\chi$.
}
\label{tab4}
\begin{tabular}{cllllll}
\hline\hline
\multicolumn{1}{c}{$L_t$}&
\multicolumn{1}{c}{$\gamma$}&
\multicolumn{1}{c}{$\sqrt{\sigma}_{T=0}$}&
\multicolumn{1}{c}{$10^4\chi_{\,T=0}$}&
\multicolumn{1}{c}{$t$}&
\multicolumn{1}{c}{$10^4\chi$}&
\multicolumn{1}{c}{$R$}\\
\hline\hline
6     & 0.335  & 0.296(2)   & 2.27(2) & $-$0.088(9) & 2.30(4) & 1.01(2) \\

      & 0.3365 & 0.279(3)   & 1.76(2) & $-$0.032(13)& 1.80(4) & 1.02(2) \\

      & 0.3369 & 0.275(2)   & 1.66(4) & $-$0.018(10)& 1.40(4) & 0.84(3) \\

      & 0.338  & 0.264(1)   & 1.416(9)& $\phantom{-}$0.022(8) & $^*$0.349(5) & $^*$0.246(4) \\

      & 0.3395 & 0.258(3)   & 1.11(3) & $\phantom{-}$0.047(14)  & $^*$0.155(3) & $^*$0.139(5) \\
 
      & 0.341  & 0.2368(6)  & 0.896(10) & $\phantom{-}$0.140(9) & $^*$0.0792(13) & $^*$0.089(2) \\
\hline
   8  & 0.344  & 0.2160(8)  & 0.608(7)  & $-$0.060(6)  & 0.607(13) & 1.00(2) \\

      & 0.346  & 0.206(2)   & 0.48(3)   & $-$0.015(11) & 0.39(2) & 0.81(7) \\

      & 0.3465 & 0.202(1)   & 0.449(15) & $\phantom{-}$0.005(7) & 0.198(14) & 0.44(3) \\

      & 0.347  & 0.1981(5)  & 0.425(10) & $\phantom{-}$0.025(6) & 0.132(8) & 0.31(2) \\

      & 0.348  & 0.1940(6)  & 0.357(15) & $\phantom{-}$0.046(6) & 0.074(3) & 0.207(12) \\
\hline\hline
\end{tabular}
}

The range of $\gamma$ values that can actually be explored with the
Wilson action is rather limited. On the one hand, $\gamma$ has to be
sufficiently large so that the system is actually in the weak-coupling
region, i.e. beyond the first-order bulk phase transition at
$\gamma=0.339$ in the case $N=6$, and beyond the crossover region
characterized by a peak of the specific heat at $\beta=0.325$ for
$N=4$; see Ref.~\cite{DelDebbio:2001kz,DelDebbio:2001sj} for a more
detailed discussion of this point.  On the other hand, as $\gamma$ is
increased, two types of difficulties arise. First, as the lattice
spacing becomes smaller, larger lattices are necessary to avoid
finite-size effects. In practice, we always try to use values of
$\gamma$ such that the spatial extent satisfies $L_s \sqrt\sigma \geq
3$, where $\sigma$ is the zero-temperature string tension, as
suggested by previous
investigations~\cite{Lucini:2001ej,DelDebbio:2001kz,DelDebbio:2001sj,Lucini:2003zr}. The
second obstacle is the increase of the autocorrelation time of the
topological modes as the continuum limit is
approached~\cite{DelDebbio:2002xa,DelDebbio:2004xh}. Such a severe
form of critical slowing down puts a stringent limit on the upper
value of $\gamma$ that can be efficiently simulated, at least with the
currently available algorithms.  Finally, since the transition is
first order and rather strong in the case $N=6$, some attention must
be paid to the dependence of the Monte Carlo results on the starting
configurations. In particular, close to the finite-temperature phase
transition, we use hot or cold starting configurations according to
the side of the transition we were investigating, to avoid hysteresis
effects.

\TABLE[ht]{
\caption{Finite-temperature data for the SU(6) gauge theory.
$\sigma_{T=0}$ and $\chi_{\,T=0}$ are respectively
the string tension and the topological
susceptibility at $T=0$ obtained using symmetric lattices.
}
\label{tab6}
\begin{tabular}{cllllll}
\hline\hline
\multicolumn{1}{c}{$L_t$}&
\multicolumn{1}{c}{$\gamma$}&
\multicolumn{1}{c}{$\sqrt{\sigma}_{T=0}$}&
\multicolumn{1}{c}{$10^4\chi_{\,T=0}$}&
\multicolumn{1}{c}{$t$}&
\multicolumn{1}{c}{$10^4\chi$}&
\multicolumn{1}{c}{$R$}\\
\hline\hline
6  &0.344  & 0.2973(5) & 1.79(6) & $-$0.071(7) & 1.83(5) & 1.02(4) \\

& 0.3444 & 0.285(2)  & 1.66(7) & $-$0.032(10)& 1.69(10) & 1.02(7) \\

& 0.3448 & 0.282(3)  & 1.50(7) & $-$0.021(13) & 1.26(7) & 0.84(6) \\

& 0.3455 & 0.268(4)  & 1.40(12)& $\phantom{-}$0.030(17) & 0.096(6)  & 0.069(7) \\

& 0.346  & 0.264(3)  & 1.29(7) & $\phantom{-}$0.045(14) & 0.061(3)  & 0.047(3) \\

& 0.348  & 0.2535(6) & 0.91(7) & $\phantom{-}$0.089(8)  & 0.0118(8) & 0.0130(13) \\
\hline\hline
\end{tabular}
}

We use the following formulas and definitions for the rescaled and
reduced temperatures:
\begin{eqnarray}
&&T_r(L_t,\gamma) 
\equiv {T \over \sqrt{\sigma}} = {1\over L_t \sqrt{\sigma(\gamma)}},\\
&& 
t(L_t,\gamma) \equiv {T_r(L_t,\gamma) - T_r(L_t,\gamma_c(L_t))\over T_r(L_t,\gamma_c(L_t))}
=  \sqrt{ {\sigma(\gamma_c(L_t))\over \sigma(\gamma)}} -1,
\label{redtemp}
\end{eqnarray}
where, as before, $\sigma$ is always computed on symmetric
lattices, i.e. at $T=0$.
In order to determine the reduced temperature according to Eq.~(\ref{redtemp}),
we used the following values of the string tension at $\gamma_c$:
$\sqrt{\sigma}=0.270(2)$ for $\gamma=0.33717$ and $N=4$ (obtained
on a $12^3\times 24$ lattice),
$\sqrt{\sigma}=0.203(1)$ for $\gamma=0.3464$ and $N=4$ (obtained
on a $16^3\times 32$ lattice), and
$\sqrt{\sigma}=0.276(2)$ for $\gamma=0.34508$ and $N=6$ 
(on a $12^3\times 24$ lattice).

The topological charge $Q$ is estimated using the cooling technique
described in Ref.~\cite{DelDebbio:2002xa}. We compute the corresponding
susceptibility $\chi\equiv \langle Q^2 \rangle /V$  and the scaling
ratio: 
\begin{eqnarray}
R(L_t,\gamma)  \equiv 
{\chi(L_t,\gamma)\over \chi(\infty,\gamma)}\, .
\end{eqnarray}
The results are reported in Tables~\ref{tab4} and \ref{tab6} 
respectively for $N=4$ and $N=6$.

Before discussing the results for the topological susceptibility, let
us assess the possible sources of systematic errors in the lattice
computation.  It is well known that topological structures may
disappear during the cooling procedure. This problem is particularly
severe at finite temperature, especially for small $L_t$, where one
may not observe clear plateaux with respect to the cooling steps, see
e.g. Ref.~\cite{DiGiacomo:1991qm}. At zero temperature, a direct
comparison with a fermionic estimator of the topological charge shows
good agreement~\cite{DelDebbio:2003rn,DGP-inprep}, supporting the idea
that the cooling method is fairly stable in this case. However, the
situation at finite temperature is more difficult to control and can
eventually generate a systematic error.  In this respect, we expect
the cooling method to perform better with increasing $N$.  In order to
minimize the possible bias due to the cooling technique, only lattices
with $L_t\geq 6$ have been considered in this study.  In
Fig.~\ref{fig:cool} we show measurements of the topological
susceptibility during cooling for $N=4$, $L_t=6,8$ and $N=6$, $L_t=6$,
and values of the couplings corresponding to the high-temperature
phase, and in particular $t\approx 0.05$.  Plateaux during cooling are
clearly observed (from 4 to 20 cooling steps) in the simulations for
$N=6$, for $N=4$ with $L_t=8$, and only in the low-temperature phase
in the case $N=4$ with $L_t=6$.  This allowed us to unambiguously
determine the topological susceptibility using the cooling method also
at finite temperature. On the other hand, for $N=4$ in the
high-temperature regime, the time-direction size $L_t=6$ was not
sufficient to provide a clear plateau, as shown in Fig.~\ref{fig:cool}
for the data corresponding to $t\approx 0.05$. Measurements based on
the cooling method become rather questionable in these cases, because
they do not provide an unambigous estimator of $\chi$. In
Table~\ref{tab4} the corresponding results are marked by an asterisk:
they were obtained after 10--12 cooling steps, but they are subject to
an uncontrolled systematic error, which probably leads to an
underestimate of $\chi$, and they should therefore be taken into
account only with due care. For small $N$ and $L_t$ other estimators
for $Q$ should be used.

\FIGURE[ht]{
\epsfig{file=cool2.eps, width=12truecm} 
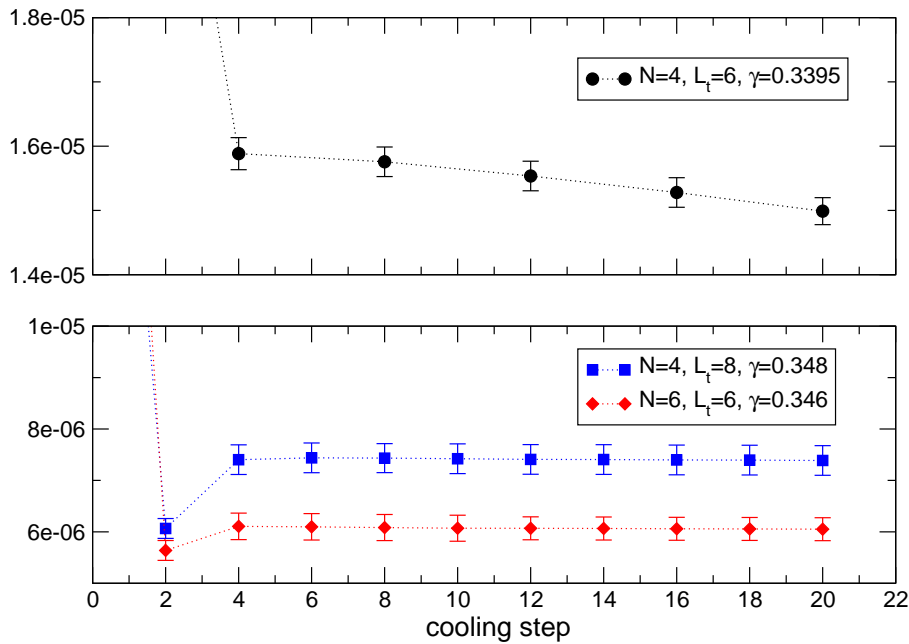
\caption{Topological susceptibility along the cooling process. The
values of $\gamma$ correspond to reduced temperatures $t\simeq 0.05$
in all cases.  }
\label{fig:cool}
}

The data for the scaling ratio $R$ are displayed in
Fig.~\ref{fig:topplot}.  One can immediately remark that its behavior
is drastically different in the low- and high-temperature phases.  In
the low-temperature phase, all data for $N=4$, $L_t=6,8$ and $N=6$,
$L_t=6$ appear to lie on the same curve, showing that scaling
corrections are small and also that the large-$N$ limit is quickly
approached.  The ratio $R$ remains constant and compatible with
$1.0$. Only close to $T_c$, i.e. for $T > 0.97 \, T_c$, does this
ratio appear to decrease. These results show that in the confined
phase the topological properties remain substantially unchanged up to
$T_c$. On the other hand, above the deconfinement phase transition, we
observe a large suppression of $\chi$. The comparison between the
$N=4$ and $N=6$ data shows that the ratio $R$ decreases much faster
for $N=6$, hinting at a vanishing large-$N$ limit of $R$ for $T>T_c$.

Numerical results supporting the same picture have also been reported
recently in Ref.~\cite{Lucini:2004yh}. The numerical evidence of the
topological suppression across the transition was inferred from
simulations at $T_c$, by monitoring the correlation of the topological
charge with the Polyakov line, whose value is used to infer the actual
phase of the configurations generated along the given Monte Carlo run.
Therefore a more quantitative comparison with our results is not
straightforward. A comparison with the results presented in
Refs.~\cite{ADD-97,Gattringer:2002mr} suggests that the suppression of
topological fluctuations is faster in SU(4) than it is in SU(3).

\FIGURE[ht]{
\epsfig{file=chit.eps, width=12truecm} 
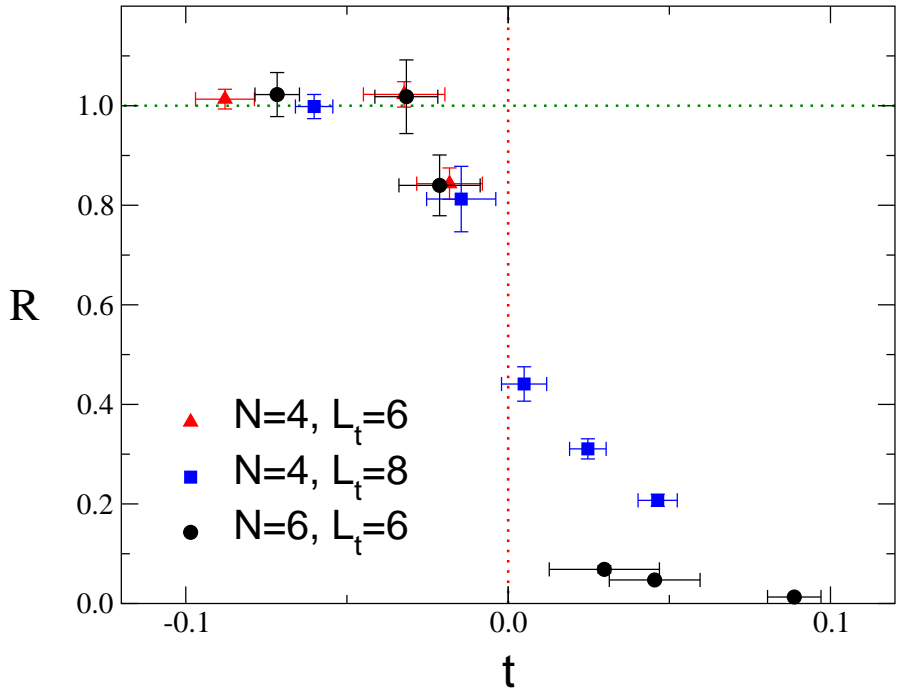
\caption{The ratio $R$ as a function of the reduced temperature $t$.
}
\label{fig:topplot}
}

A nonvanishing topological susceptibility $\chi$ implies a nontrivial 
dependence on the $\theta$ term that appears in the euclidean Lagrangian as
\begin{equation}
{\cal L}_\theta  = {1\over 4} F_{\mu\nu}^a(x)F_{\mu\nu}^a(x) - i \theta q(x)
\label{lagrangian}
\end{equation}
where $q(x)$ is the topological charge density.  Indeed $\chi$
is the second derivative of the free-energy density (ground-state energy)
$F(\theta)$ with respect to $\theta$ at $\theta=0$.  More generally,
expanding the free-energy density around $\theta=0$, one may write
\begin{equation}
F(\theta) = \frac{1}{2} \chi \theta^2 \left( 1 + b_2 \theta^2 + b_{4} \theta^4 + ... \right)
\label{fth}
\end{equation}
The parameters of the expansion of $F(\theta)$ are related to 
the moments of the probability distribution $P(Q)$ of the topological charge $Q$
in the large-volume limit.
While $\chi$  is determined from the second moment $\langle Q^2 \rangle$,
the coefficients $b_{2i}$ are related to higher moments of $P(Q)$,
for example 
\begin{equation}
b_2 = - {\chi_4\over 12 \chi}, \qquad \chi_4 = {1\over V} \left[ 
\langle Q^4 \rangle_{\theta=0} - 3 \left( 
\langle Q^2 \rangle_{\theta=0} \right)^2 \right]. 
\label{b2}
\end{equation}
The large-volume limit of the probability distribution $P(Q)$ is
Gaussian only if $b_{2i}=0$, i.e. when $F(\theta) = \frac{1}{2} \chi
\theta^2$ without higher-order corrections.  A nontrivial expansion
around $\theta=0$, such as Eq.~(\ref{fth}), reflects deviations from a
simple Gaussian behavior of $P(Q)$, whose size is controlled by the
coefficients $b_{2i}$.~\footnote{ An apparently contradictory result
has been reported in Refs.~\cite{GLWW-03,BCNW-03} for the expected
large-volume probability distribution $P(Q)$, i.e.  $P(Q)=(2\pi\langle
Q^2\rangle)^{-1/2}e^{-{Q^2\over 2\langle Q^2\rangle}}\left[1
+O(1/V)\right]$, which was obtained starting from a generic expansion
of $F(\theta)$ at $\theta=0$, such as Eq.~(\ref{fth}), and evaluating
the large-volume behavior of $P(Q) = \int d\theta \,\exp[ - i Q \theta
- V F(\theta) ]$ using a saddle point approximation. A Gaussian
behavior in the large-volume limit~\cite{GLWW-03} would contradict the
assumption of a generic expansion of $F(\theta)$, and, in particular,
a nonzero value of $b_2$, which implies a nonzero fourth moment of the
large-volume $P(Q)$.  The point is that the contributions considered
as $O(1/V)$ corrections to the Gaussian behavior cannot be neglected
in order to reproduce the correct physically relevant large-volume
limit of the distribution's moments. This can be checked by computing
the corrections to the saddle-point approximation.  
} The coefficient $b_2$ has been
estimated in Ref.~\cite{DelDebbio:2002xa} for the SU($N$) gauge
theories with $N=3,4,6$ at $T=0$, finding very small values,
i.e. $b_2=-0.023(7)$ for $N=3$, $b_2=-0.013(7)$ for $N=4$, and
$b_2=-0.01(2)$ for $N=6$, supporting the conjecture
\cite{Witten-98,Gabadadze-99} $b_2=O(1/N^2)$.  Thus for $N\ge 3$ the
simple Gaussian form $F(\theta)\approx \frac{1}{2} \chi \theta^2$ is
expected to provide a good approximation of the dependence on $\theta$
for a relatively large range of values of $\theta$.  In order to
investigate this issue at finite temperature, we have also computed
$b_2$ in our finite-temperature Monte Carlo simulations.  In the
low-temperature phase the estimates of $b_2$ turn out to be compatible
with those at $T=0$, suggesting that $F(\theta)$ remains substantially
unchanged up to $T=T_c$, with very small corrections to the Gaussian
behavior.  On the other hand, in the high-temperature phase the
absolute value of $b_2$ turns out to be significantly larger, for
example at $t\simeq 0.05$ we found $b_2\simeq -0.05$ for $N=4$ and
$b_2\simeq -0.08$ for $N=6$, indicating larger deviations from the
Gaussian behavior, although they are still moderately small.

In conclusion, the results presented in this paper suggest that the physical
properties determined by the fluctuations of the topological charge,
such as the Witten--Veneziano relation (\ref{eq:WV}), remain
substantially unchanged in the low-temperature confined phase.  In
particular, the topological susceptibility of the pure gauge theory
has a nonvanishing large-$N$ limit in the low-temperature phase, as at
$T=0$.  On the other hand, in the high-temperature phase there is a
sharp change of regime where the topological susceptibility is largely
suppressed.  Such suppression becomes larger with increasing $N$,
suggesting that the topological charge vanishes above the critical
temperature. Monte Carlo results seem to support the scenario
presented in~\cite{KPT-98}: at large $N$ the topological properties in
the high-temperature phase, for $T>T_c$, are essentially determined by
instantons from very high temperature down to $T_c$; the exponential
suppression of instantons induces the rapid decrease of the
topological activity observed in the large-$N$ limit.

\acknowledgments{ We 
greatly benefitted from discussions with M. Campostrini,
L. Giusti, M. Mintchev, A. Pelissetto  and 
G. Veneziano. LDD thanks the Physics Department of the
University of Pisa for kind hospitality during the final stages of
this work. 
We are also indebted to Maurizio Davini for his valuable
technical support.
}

\end{document}